\documentclass[aps,prl,twocolumn,showkeys,showpacs]{revtex4}
\usepackage{graphicx}
\usepackage{natbib}

\begin{document}

\newcommand{\FU}{\affiliation{Freie Universit\"at Berlin, Institut f\"ur Experimentalphysik, Arnimallee 14, 14195 Berlin,
Germany}}
\newcommand{\Beff}{\ensuremath{B_{e\!f\!f}}}

\title{Realtime Magnetic Field Sensing and Imaging using a Single Spin in Diamond}

\author{Rolf Simon \surname{Schoenfeld}}
\author{Wolfgang \surname{Harneit}}\email{w.harneit@fu-berlin.de}
\FU

\date{\today}%
\newcommand\revtex{REV\TeX}
\keywords{MAGNETIC RESONANCE; SINGLE SPIN; OPTICAL MICROSCOPY; DIAMOND; SUPERRESOLUTION; NITROGEN-VACANCY CENTERS; MAGNETIC IMAGING}

\begin{abstract}

The Zeeman splitting of a localized single spin can be used to
construct a magnetometer allowing high precision measurements of
magnetic fields with almost atomic spatial resolution. While
sub-$\mu$T sensitivity can in principle be obtained using pulsed
techniques and long measurement times, a fast and easy-to-use method
without laborious data post-processing is desirable for a
scanning-probe approach with high spatial resolution.
In order to
measure the resonance frequency in realtime, we applied a
field-frequency lock 
to the continuous wave ODMR signal of a single electron spin in a
nanodiamond. In our experiment, we achieved a sampling rate of up to
100 readings per second with a sensitivity of 6
$\mu$T/$\sqrt{\textrm{Hz}}$. Using this method we have imaged the
microscopic field distribution around a magnetic wire. Images with
$\sim 30\, \mu$T resolution and 4096 sub-micron sized pixels were
acquired in 10 minutes. By measuring the field response of multiple
spins on the same object we were able to partly reconstruct the
orientation of the field.

\end{abstract}

\maketitle

The nitrogen vacancy defect in diamond (NV center) has been recently
proposed as a magnetic field probe with the potential for
extraordinary spatial resolution, field precision, linearity and
directional sensitivity \cite{Degen2008apl, Balasubramanian2008,
Maze2008}. Its outstanding physical properties allow the spin state
of a single center to be read out via optically detected magnetic
resonance (ODMR) using confocal techniques \cite{Gruber1997}. This
opens promising possibilities in magnetic imaging
\cite{Chernobrod2005} with performances comparable to SQUIDs
\cite{BLACK1993, Finkler2010}, magnetic-resonance force microscopy
(MRFM) \cite{Mamin2007, Rugar2004} and similar high-resolution
attempts \cite{Blank2009}, with the added benefit of
room-temperature operation. In contrast to the widely used magnetic
force microscopy (MFM \cite{Martin1987}), single spin magnetometry
can be made independent of magnetic field gradients.

Scanning-probe concepts using a single NV center \cite{Taylor2008, Cole2009, Cappellaro2009, Hall2009} can be distinguished from multiplexing approaches where many NV centers are read out simultaneously \cite{Taylor2008,
Maertz2010, Steinert2010}. While the latter approach offers fast acquisition and full reconstruction of the magnetic field vectors, it is inherently limited in field resolution by dipolar interaction between the NV centers and in spatial resolution by the optical diffraction limit. Moreover, time-consuming post-processing of the data is necessary. Single-center scanning probes, on the other hand, provide significantly higher resolution in field and space, but this approach is in general considered too slow for imaging \cite{Steinert2010}. Indeed, although contour lines corresponding to a single constant magnetic field were made visible in scanning confocal fluorescence images \cite{Balasubramanian2008}, true field images were so far obtained only with the multi-center approach \cite{Maertz2010, Steinert2010}. In all of these imaging-related studies, the full ODMR spectrum was acquired for each pixel using continuous (cw) irradiation.

Alternate ways to magnetic field sensing based on pulsed excitation
and detection schemes have been proposed but not yet used for
imaging, although experiments have demonstrated the soundness of the
sensing principles \cite{Maze2008, Balasubramanian2009, Shin2010}.
While echo-based methods provide a better field sensitivity, they
are only sensitive to periodically alternating fields. Static
magnetic structures can be imaged only if the NV spin is mounted on an oscillating tip. Then, similar
to the widely used MFM technique \cite{Martin1987}, only strong field
gradients can be measured. For weaker gradients, those methods are less suited and Ramsey
fringes can be used instead. In all of these pulsed-ESR methods,
pulsed light excitation and detection is required in addition to
pulsed microwave channels.

\begin{figure}
\includegraphics[scale=0.82]{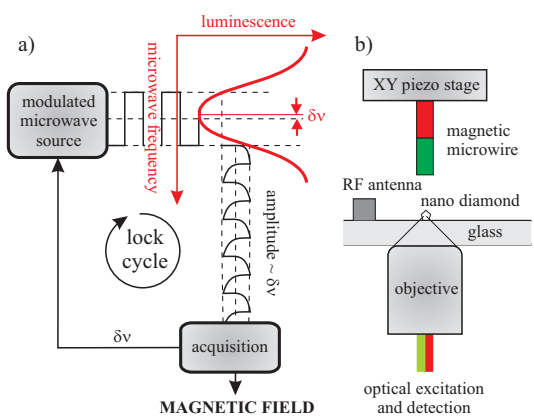}%
\caption{(color online) (a) A single NV spin responds to a
frequency-modulated microwave close to resonance with a luminescence
modulation proportional to the frequency offset. The luminescence is
recorded via lock-in principle and used for frequency correction and
local magnetic field calculation. a) A 100 $\mu$m magnetized steel
wire is scanned over a nano-diamond containing a NV center. The
microwave is generated by an antenna $\sim100\,\mu$m
apart.}\label{fig:principle}
\end{figure}

In this letter, we present an alternative experimental approach to
measure static fields with a single spin scanning probe, which uses
only continuous optical excitation and frequency-modulated
microwaves and is thus far less demanding in resources. Our approach
offers a sensitivity comparable with the related pulsed technique
(Ramsey fringes) and can be seamlessly integrated as a "tracking and
surveying" mode with the echo approaches if higher sensitivities are
needed.

It is easy to see from the Hamiltonian of the diamond electron spin $(S = 1)$
\begin{equation}\label{eq:NVham}
H = \gamma \Beff S_z + D {S_z}^2 + E\left(S_x^2-S_y^2\right)
\end{equation}
that for an exact determination of the local field strength \Beff\
along the symmetry axis of the NV center, only knowledge of the
resonance position of one of the two allowed electron spin
transitions is necessary as long as the zero field spectrum of the
center is known. Here, $\gamma$ is the gyromagnetic ratio of the
spin and $D$ and $E$ are its zero field splitting parameters. In
order to constantly track the resonance position without measurement
of the complete spectrum, we use a modification of the
field-frequency lock (Fig.~\ref{fig:principle}(a)) known from
magnetic resonance \cite{MACIEL1967}.

\begin{figure}
\includegraphics{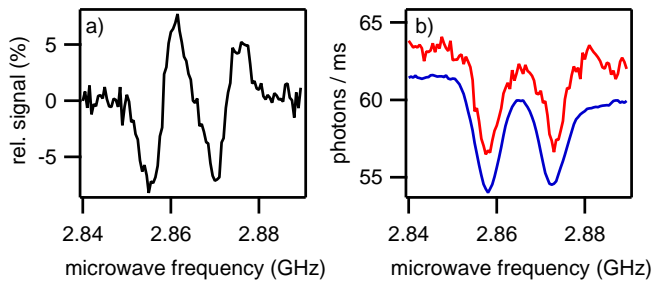}%
\caption{ODMR spectrum of a NV center with $E \neq 0$. (a) FM
detected signal, $r_{mod}= 100 \textrm{ kHz}$, $\nu_{mod} = 2.5
\textrm{ MHz}$\qquad (b) integral of (a) (lower trace), CW signal
(upper trace) shown for comparison. The sampling time was 100 ms per
point for all measurements.}\label{fig:FM_vs_CW}
\end{figure}

We use a rectangular frequency modulation (FM) of the microwave,
created using two alternately switched and equally leveled signal
sources
\begin{equation}\label{eq:FM}
\nu(t) = \bar{\nu} + \nu_{mod} \Sigma\left[\cos(2\pi r_{mod} t)\right],
\end{equation}
where $\bar{\nu}$ is the central frequency, $\nu_{mod}$
 the modulation amplitude, $r_{mod}$ the modulation rate, and $\Sigma$ is
the sign function. In our detection, we use a software lock-in
resulting in a signal proportional to the first derivative of the
spectrum $L(\nu)$,
\begin{equation}\label{eq:lockin}
S(\bar{\nu}) \propto 2\nu_{mod} \frac{\partial L}{\partial\bar{\nu}}
\end{equation}
as long as the modulation amplitude is small compared to the width
of the peak (Fig.~\ref{fig:FM_vs_CW}(a)).
The original spectrum can be obtained by integration resulting in a
smoother curve compared to a straightforward CW measurement
(Fig.~\ref{fig:FM_vs_CW}(b)). In case of unbiased photon shot noise
this corresponds to an increased signal-to-noise ratio (SNR). One
should note that the higher SNR is accompanied by a loss of
resolution, which is limited by $\nu_{mod}$.

\begin{figure}
\includegraphics{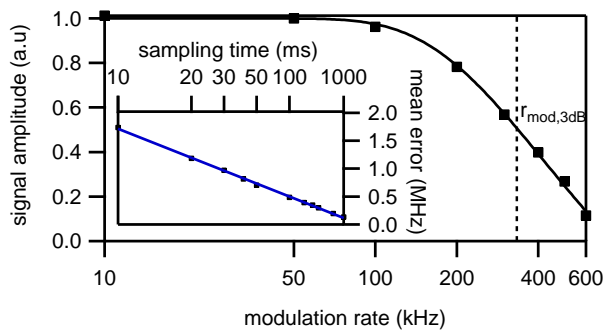}%
\caption{Amplitude dependence of the modulation detected signal on
the modulation rate, showing a response time $\tau_{3dB} =
1/2r_{mod,3dB} = 1.5\,\mu$s ($P_{RF} = 1\,$W, $P_{Laser} = 50\,\mu$W
at objective entry). The inset shows the mean error in the resonance
frequency derived from the signal after averaging multiple
modulation cycles. The results are in excellent agreement with a
photon shot noise of 170 kHz
$/\sqrt{\textrm{Hz}}$.}\label{fig:performance}
\end{figure}

The speed of our technique relies on a fast modulation of the ODMR signal. One of the great advantages of the NV center is that its luminescence is not only spin-dependent but is also able to follow a change %
in $|S_z|$
within less than a microsecond. This is due to the fact that
moderate laser illumination creates high spin polarization in a
microsecond while the typical inverse Rabi nutation frequency is in
the same order of magnitude. For simplicity, in this study we
consider a phase relaxation time $T_2$ of a few microseconds (as is
the case for our type Ib nanodiamonds), so that coherent effects can
be neglected. This allows us to operate at $r_{mod} = 100$ kHz, a
value common in conventional electron spin resonance. The dependence
of the signal amplitude $|S|$ on $r_{mod}$ is shown in
Fig.~\ref{fig:performance}, showing that the practical limitation is
given by the repolarization rate. A response time $\tau_{3dB} \simeq
1.5\,\mu$s can be estimated based on the criterion of a 3 dB
attenuation of the signal.

Applying Eq.~(\ref{eq:lockin}) to a gaussian peak, it follows that a
signal measured for a fixed microwave center frequency, but slightly
off resonance, is proportional to the frequency offset. The true
resonance position, and hence the magnetic field via
Eq.~\ref{eq:NVham}, can be calculated. It is thus possible with our
approach to locate and follow the resonance position of the NV
center precisely with millisecond reaction time provided that the resonance position
does not change more than the FWHM of the ODMR peak between two
regulation cycles. A realtime field measurement can be conducted by
continuously adjusting and recording the microwave frequency using
this signal.

We now discuss the precision and speed of our method in terms of
sensitivity $\eta = \delta B \sqrt{T}$, defined by the field error
$\delta B$ at given measurement time $T$. For a standard deviation
$\sigma$ of the measurement and a general repetition rate $r$ (here:
$r = 2r_{mod}$, in pulsed techniques, $r=1/t$ with $t$ =
interrogation time, e.g., the spacing of Ramsey $\pi/2$ pulses) one
obtains for any method in the photon shot noise limit
\cite{Taylor2008}
\begin{equation}\label{eq:sensitivity}
\eta = \frac{\sigma}{\sqrt{r} dS/dB}
\end{equation}
where $dS/dB$ is the signal response to a change in the magnetic
field. For d.c.~measurements, a limit $r = 1/T_2^\ast$ obtains.
Considering similar signal contrast and noise level at equal
experimental conditions, and neglecting the photon collection duty
cycle which is one in our case and close to one using pulsed methods
at high repetition rate, any difference in sensitivity can be
attributed to $dS/dB$. For the Ramsey fringe method, $dS/dB =
\gamma\,T_2^\ast$ \cite{Taylor2008}. In our approach, assuming an
inhomogeneously broadened line $S(\omega)=\exp\left[-\ln2
\left(\omega T_2^*\right)^2\right]$ and an optimal modulation
amplitude $\nu_{mod} = \sqrt{2/\ln2}/T_2^\ast$, we find
\begin{equation}
\frac{dS}{dB} = \sqrt{\frac{8 \ln2}{e}} \;\gamma T_2^\ast \approx
1.4 \;\gamma T_2^\ast.
\end{equation}

We conclude that our method shows comparable and even slightly
improved performance compared to Ramsey fringes, leading to a
theoretical sensitivity better than 1 $\mu$T$/\sqrt{\textrm{Hz}}$ (see \cite{Taylor2008} for details).
Our measured sensitivity shown in Fig.~\ref{fig:performance} is
slightly reduced due to experimental issues, i.e., lower signal
contrast due to weaker microwave field and optical broadening due to
high laser power. The latter can be avoided using an alternating
optical/microwave excitation pattern. In this study, we decided to
keep the implementation as simple as possible in order to emphasize
a potential advantage of our method; although we used two
pulse-switched microwave generators and time-triggered detection for
convenience, an easier technical implementation is possible using
only a single modulated microwave source and phase-synchronized
detection of a diode current.

\begin{figure}
\end{figure}
\begin{figure}
\includegraphics{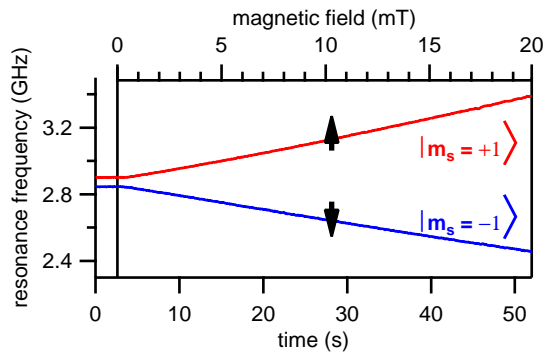}%
\caption{Spin transition frequencies vs.~external field strength
measured by tracking the resonance positions during a magnetic field
sweep (sampling rate 25 Hz).}\label{fig:breitrabi}
\end{figure}
\begin{figure}
\includegraphics{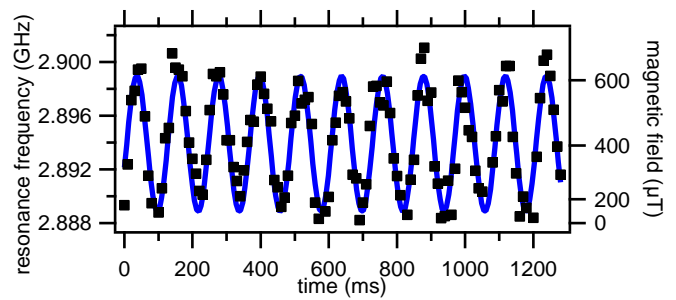}%
\caption{Realtime measurement of an ac magnetic field. $f_{ac}$
 was 8.3 Hz corresponding to a maximum magnetic
field gradient of $\sim$ 10 mT/s (sampling rate 100
Hz).}\label{fig:fasttrack}
\end{figure}

\begin{figure*}
\includegraphics[scale=0.9]{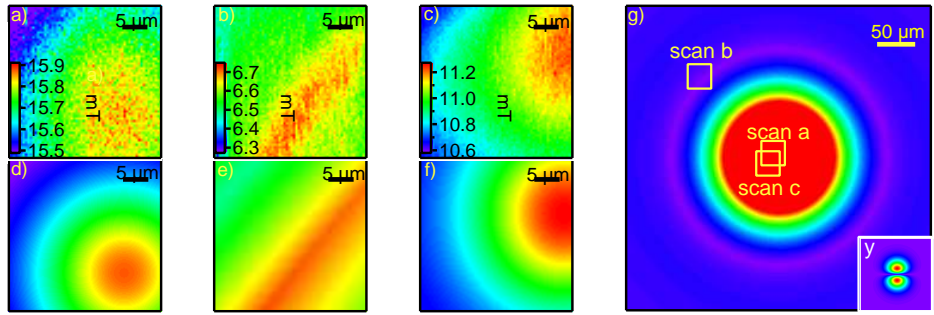}%
\caption{Magnetic field images taken with single NV centers. (a) NV
axis normal to xy-plane ($\theta = 0^\circ$), (b) NV axis projection
on xy-plane along x, pointing slightly downwards ($\phi = 0^\circ$,
$\theta = 100^\circ$), (c) $\phi = -60^\circ$ (clockwise), $\theta =
45^\circ$, (d)-(f) simulation of (a)-(c) assuming a simple dipole
model (Eq.~\ref{eq:dipole}), no $B_xS_x$ components have been taken
into account, (g) simulation of the normal magnetic field showing
the scan areas, the inset shows an in-plane field component.}
\label{fig:fieldImages}
\end{figure*}

A realtime resonance tracking of both transitions during an
uncalibrated independently driven magnetic field ramp created by an
external magnet is shown in Fig.~\ref{fig:breitrabi}. The result is
in perfect agreement with the Zeeman splitting expected and shows
the excellent linearity of the probe in a broad field range. Note
that in our experiment that range (2.4 to 3.4 GHz) it is given by
the bandwidth of the involved microwave components. For tracking we employed a
locking algorithm in which $S(\nu)$ is measured for a time $T$ and
the microwave center frequency is adjusted by the factor calculated
from the modulated spectrum measured at zero field. Locking of the
resonance position is achieved as long as the frequency changes
between two measurements are smaller than the peak width. Faster
changes will eventually lead to no further correction since the
signal derivative is zero both at the peak maximum and far away from
it (at the baseline). This marks the breakdown of the lock. Yet, as
long as the peak is still locked the mechanism now enables us to
track the resonance position over time. The response time is
ultimately limited by the photon output and signal contrast of the
color center. Taking typical values for a NV center (200 photons per
ms, signal contrast $\alpha = 0.2$), the ultimate theoretical limit
is $\sim0.2$ ms, because $\sqrt2/\alpha^2$ photons are needed for a
SNR of 1.

For a speed test simulating imaging conditions we applied an ac
magnetic field with variable frequency $f_{ac}$ (Fig.~
\ref{fig:fasttrack}). $f_{ac}$ was set to the maximum value where
the probe could still follow the field without interruptions due to
lock break-off and can be used to estimate the maximum line
frequency of a scanning probe measurement. Higher sampling rates
than the one used here will lead to even higher scanning
frequencies, but here we were limited by the dead time of the
counter card used for fluorescence readout.

Having now determined speed, range, and sensitivity of our probe, we
measured the stray magnetic field of a 100 $\mu$m diameter steel
wire magnetized along its axis. The wire was mounted on a XY
piezoscanner placed above the sample containing nanodiamonds. The
distance to the sample was about 100 $\mu$m. The scan was taken
using an integration time of 150 ms per point. Due to the limited
range of the piezo-scanner, only small areas of the whole field were
measured. We measured several points of interest using different
centers with known axis direction (Fig.~\ref{fig:fieldImages}). Each
measurement took about 20 minutes for two 64x64 pixel images (back-
and forth scan direction). No external field was applied during the
experiment. The NV center axes were determined via realtime field
measurements while rotating an external magnet; this was done in
absence of the sample in order to avoid changes in magnetization.
Assuming the wire to behave like a magnetic point dipole at this
distance, we simulated its magnetic field using
\begin{equation}\label{eq:dipole}
\vec{B} =
\frac{\mu_0}{4\pi}\frac{3\vec{r}\left(\vec{m}\cdot\vec{r}\right)-\vec{m}r^2}{r^5}\qquad.
\end{equation}
The measured areas showed excellent agreement to the simulation. 
Comparing results from different centers of the same area we could
find good agreement of the overall strength of the measured fields.
For example by comparing the field maximum in scan (a) and (c) one
finds that $B_z = B(\textrm{scan a}) = B(\textrm{scan c}) /
\cos{(\theta_c = 45^\circ)}$, thus confirming the direction of the
magnetic field vectors below the wire pointing perfectly
perpendicular to the plane.

In conclusion, we have demonstrated a robust and fast responding
implementation of a magnetic field measurement using a single spin
in diamond. We have shown that a single spin in diamond is well
suited for use in a scanning probe microscope and that also
directional information can be obtained with little effort. Our approach will allow to take precise 2D field
images with nanometer resolution at ambient conditions within a few
minutes which might open up fascinating new insights in cell
biology, nanoelectronics and magnetic nanostructures. The only issue
remaining for substantial local resolution increase is a better
distance control to the sample, e.g. reliable fabrication of
nanodiamonds on an AFM tip, which is currently under development
\cite{Sar2009, Ampem-Lassen2009, Cuche2010}.

This work was supported by the Volkswagenstiftung through the program "Integration of molecular components in functional macroscopic systems".


\end{document}